\documentclass[12pt, a4paper]{article}
\usepackage{graphicx}
\usepackage{longtable}
\usepackage{array}
\usepackage{calc}

\newcommand{\address}[1]{\begin{itemize}
   \item[]\rm\raggedright #1
   \end{itemize}}

\begin{document}
\bibliographystyle{acta}

\title{Computer simulation of the microstructure and rheology of semi-solid alloys under
    shear}
\author{Michel Perez\,$^\dag$\ \footnote{email: Michel.Perez@gpm2.inpg.fr}\ , Jean-Charles
Barb\'e\,$^\ddag$\ , Zoltan Neda\,$^\S$\ , \\ Yves
Br\'echet\,$^\P$\ and Luc Salvo\,$^\dag$}

\maketitle

\address{$^\dag$\ Laboratoire G\'enie Physique et M\'ecanique
des Mat\'eriaux,\\ UMR CNRS 5010, BP 46, 38 402~Saint Martin
d'H\`eres Cedex, France}

\address{$^\ddag$\ Laboratoire de Solidification et de ses
Proc\'ed\'es,\\ DTA/SPCM, CEA Grenoble, 17 rue des Martyrs, 38
054~Grenoble Cedex 9, France}

\address{$^\S$\ Babes-Bolyai University, Faculty of Physics, Dept. of
Theoretical Physics, RO-3400 Cluj-Napoca, Romania}

\address{$^\P$\ Laboratoire de Thermodynamique et de Physico-Chimie
M\'etallurgique,\\ UMR CNRS 5614, BP 75, 38 402 Saint Martin
d'H\`eres Cedex, France}



\begin{abstract}
The rheological behavior of metallic alloys containing both solid
and liquid phases is investigated in the low solid fraction range
($<$50\%). This behavior depends on both the solid fraction and
the shear rate. The concept of Effective Volume Fraction (EVF) is
used to decorrelate the influence of these two parameters. At high
shear rate the slurry behaves like a suspension of hard spheres,
whereas at lower shear rate, particles tend to aggregate in
clusters, entrapping liquid and thus, increasing the EVF and the
viscosity. A lattice model is introduced to simulate the
aggregation / break-up processes within a slurry under shear. When
the steady state is reached, the entrapped liquid fraction is
calculated, leading to a viscosity estimation. Simulation results
for the viscosity and 3D cluster structure are in good agreement
with experimental results.
\end{abstract}

\section{Introduction}

Semi-solid slurries are characterized by the coexistence of solid
and liquid phases. They are usually observed in alloys with two or
more constituents for temperatures between the solidus and
liquidus lines. The rheological properties of this dual phase
state is of interest both for casting and for metal forming
operations known as thixoforming \cite{Flemings91}. When submitted
to shear, the steady state viscosity decreases as the applied
shear rate increases, reaching values of order 100~mPa.s
\cite{Joly76}.

This behavior is called rheofluidization and it is usually
explained by the interaction between the solid particles. It is a
very general phenomenon observed in suspensions, metalic alloys in
the semi-solid state, colloids, latexes, {\it etc}. Their
characteristic behavior is qualitatively interpreted as a
competition between the aggregating (Coulomb attraction, Wan Der
Waals forces, surface forces...) and break-up forces (shear). At
low shear rate (typically lower than 1~s$^{-1}$), individual
particles can aggregate in "clusters", being able to form a more
or less rigid network and the slurry is considered as a solid. At
very high shear rate, the motion of the particles prevents the
particle/particle bonding and leads to a more dispersed suspension
behaving more like a fluid.

The concept of effective volume fraction introduced by Quemada
\cite{Quemada85} allows to relate the cluster characteristics to
the viscosity of the mixture. The aim of the present paper is to
build on this idea by considering the dynamics of cluster
formation, and to derive a phenomenology for non-newtonian
behaviour of semi-solid slurry, which is based on an understanding
of the elementary phenomena governing the cluster dynamics.

\begin{figure}
\begin{center}
\includegraphics[width=\textwidth]{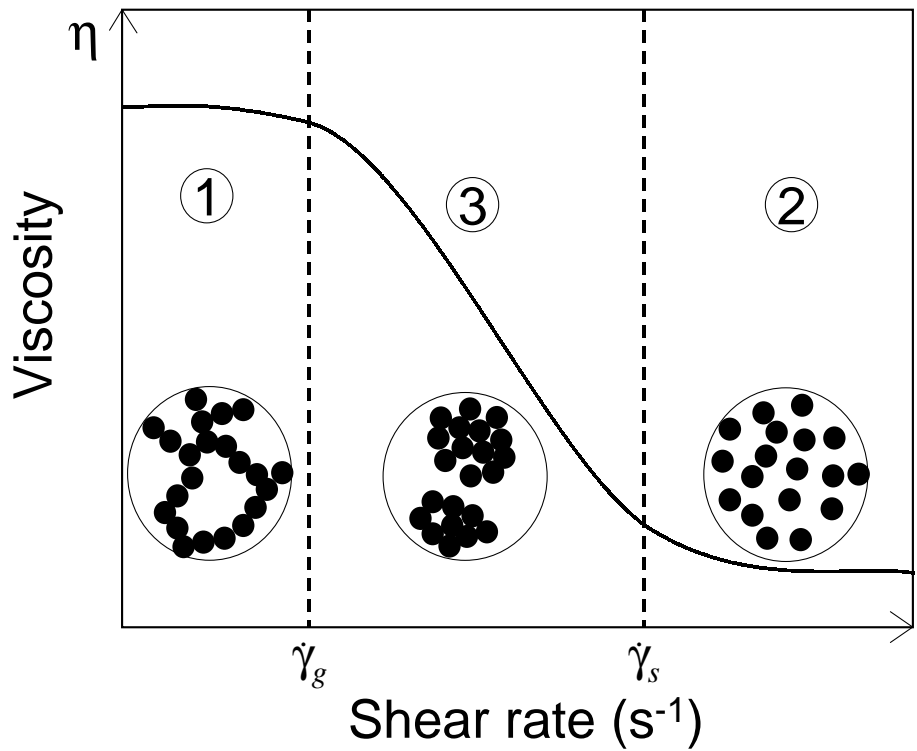}
\caption{Rheofluidization: Viscosity decreases between the shear
rates $\dot{\gamma}_g$ and $\dot{\gamma}_s$. 1- Percolating
network, 2- dispersed suspension, 3- suspension of clusters.}
\label{rheofig}
\end{center}
\end{figure}

The rheofluidization has been extensively studied (see for
instance the work of Quemada \cite{Quemada98}). Figure
\ref{rheofig} shows a typical shear rate dependence of the
viscosity for an interacting particles suspension. Shear
thickening that may occur for higher values of the shear rate
\cite{Barnes89} is not considered. The behavior at high shear rate
when particles are well separated has been accurately described by
Krieger \cite{Krieger59}:

\begin{equation}\label{kriegereq}
  \eta = \eta_0 \left( 1- \frac{\Phi}{\Phi_M} \right)
  ^{-2.5 \Phi_M}
\end{equation}

$\Phi$ being the solid fraction of the suspension and $\Phi_M$ its
maximum value (close packing). Below a given value
$\dot{\gamma}_s$, an increase of the viscosity is observed. This
can be interpreted as an increase in  the effective volume
fraction $\Phi_{eff}$, which takes into account both the solid and
the entrapped liquid. $\Phi_{eff}$ replaces $\Phi$ in
equation~\ref{kriegereq}.

Jeffrey and Acrivos \cite{Jeffrey76} underlined the notion of
structure resulting from aggregation between solid particles: for
the same solid fraction $\Phi$, different structure can lead to
different viscosities. The structure depends itself on the shear
rate. This statement is easily understandable for the two extreme
limits of the shear rate: "percolating network" for low shear rate
and "disperse suspension" for high shear rate. Between these two
domains, the structure of diphasic solutions is not fully
understood due to several reasons:
\begin{itemize}
\item Mean-field modeling fails to describe the dynamics of such
slurries because it does not take into account the interaction
between solid particles.
\item 2D observations are useful but not sufficient because they
do not show the real state of aggregation. In particular, 3D
connectivity of the solid particles cannot be easily revealed.
Serial cutting \cite{Ito92} allows 3D reconstruction, but is only
tractable for relatively small zones.
\item Difficulty of 3D observations: Small Angle Scattering is
efficiently used \cite{poulain99} with colloids, but gives only a
characteristic length-scale of the suspension. Moreover, it is
limited to small solid particles (less than a few microns). A new
observational technique is under development:
synchrotron-radiation microtomography allowing 3D investigation by
means of phase contrast of a relatively large part of a sample
with a resolution of a few micrometers \cite{Buffière99}.
\end{itemize}

In a previous work \cite{Barbé00}, an analytical estimation of the
characteristic radius of clusters was derived considering that
aggregation and coalescence lead to spherical compact clusters.
This radius was found to decrease with the shear rate as
$\dot{\gamma}^{-4/7}$. In this paper we discuss the aggregation
and break-up phenomena leading to more or less opened structures
neglecting, this time, the densification of the resulting
clusters.

Different approaches has been used to simulate the rheological
behaviour of colloidal suspensions. Dynamic simulation tools such
as nonequilibrium Brownian dynamics \cite{Rastogi96} and Stokesian
dynamics \cite{Bradi88} have provided insights into the influence
of the various colloidal forces on the microstructure and the
effect of this microstructure on suspension rheology.
Non-dimensional structural parameter has been introduced to
predict the rheology of aggregated sediment suspension
\cite{Toorman97} and semi-solid slurries \cite{Martin94}.

Based on simple physical assumptions, we propose in this paper an
original computer simulation of the microstructure of a semi-solid
metallic alloy submitted to shear. First, we discuss the
parameters governing the structure, then we use a simple approach
to relate the microstructure with rheological properties.

\section{Structure and viscosity of sheared
suspension}\label{rheosec}

The role of the solid fraction, $\Phi$, in the rheology of sheared
suspensions has been extensively studied (see for instance the
review of Rutgers \cite{Rutgers62}). The most used law for the
$\Phi$ dependence of the viscosity is the Krieger phenomenological
law. It has the advantage of pointing out the concept of maximum
packing fraction $\phi_M$.

\begin{figure}
  \centering
  \includegraphics[width=\textwidth]{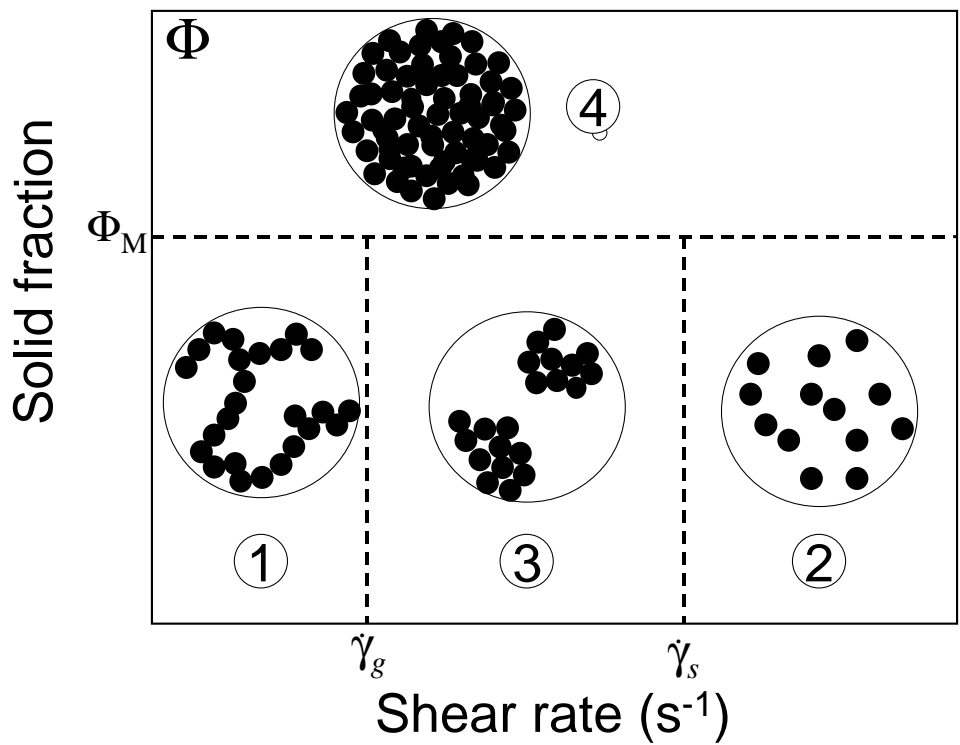}
  \caption{Structure of the  suspension in the ($\dot{\gamma}$,
  $\Phi$) phase-space. 1-Gel, 2-Suspension of individual particles,
3-Suspension of clusters, 4-Compact arrangement. In this paper we
are mostly interested in regions 2 and 3.}\label{structurefig}
\end{figure}

The role of the shear rate is less clear. We know that it tends to
break-up the clusters: depending on the applied shear rate and the
solid fraction. The possible structures can be schematically
grouped in four different classes (see figure~\ref{structurefig}):

\begin{itemize}
\item At low solid fraction ($\Phi \leq \Phi_M$), the structure depends on the shear rate.
At low shear rate (1), very little break-up occurs, leading to a
3D interconnected network. At high shear rate (2), the break-up
process dominates leaving the system as a dispersed suspension
(individual particles are well separated). In between those two
domains (3), the aggregation and break-up processes
counterbalance, leading to a suspension of clusters with
$\dot{\gamma}$ dependent characteristic size and shape.
\item At high solid fraction ($\Phi \geq \Phi_M$), close-packing
is observed for any applied shear rate. The shear is then
localized in particular planes.
\end{itemize}

The frontiers delimiting the different domains in
figure~\ref{structurefig} should be seen more as transition zones
than strict boundaries.

In the dispersed suspension region the viscosity can be accurately
described with a model developed for suspensions of hard spheres
(see equation \ref{kriegereq}). In the 3D interconnected network
region the deformation mechanisms are totally different involving
much more the solid than the liquid. Pseudo plastic models leading
to a power law, $\eta = m \dot{\gamma}^{n-1}$, describes
successfully the rheology of the two-phased material (see
reference\cite{Laxmanan80}) .

In the intermediate region, Quemada \cite{Quemada85} introduced
the concept of Effective Volume Fraction (EVF, $\Phi_{eff}$ in
equation \ref{kriegermod}) to replace the solid fraction $\Phi$ of
equation \ref{kriegereq}, turning it into:

\begin{equation} \label{kriegermod}
  \eta = \eta_0 \left( 1- \frac{\Phi_{eff}}{\Phi_M} \right)
  ^{-2.5 \Phi_M}
\end{equation}

The EVF takes into account the entrapped liquid not involved in
the hydrodynamic flow. $\Phi_{eff}$ is then the sum of the real
solid fraction $\Phi$ and the part of the liquid which does no
longer take part in the liquid flow. It could be, for example the
entrapped liquid in the middle of a solid aggregate. Note that
liquid does not need to be completely surrounded by solid to be
entrapped (see section \ref{simulsec}). On this basis the
structure of the suspension is re-injected into equation
\ref{kriegereq} through its correlation with the EVF. For a
suspension of compact clusters, the EVF would be equal to the
solid fraction, while for a more open structure (containing more
entrapped liquid) it could be much higher than $\Phi$ leading to
higher values of viscosity. The maximum packing fraction $\Phi_M$
characterizes a geometrical compactness. It is a function of the
size and shape distribution of the individual solid particles and
does not depend on $\dot{\gamma}$. The EVF takes into account the
dependence of the microstructure as a function of the applied
shear rate. The structure could have been introduced in a shear
dependent packing fraction $\Phi_M$ \cite{Wildemuth84}, instead of
$\Phi$. Both approaches lead to the same results.

Ito and Flemings \cite{Ito92} used this concept to interpret
experimental results on semi-solid slurries. After a time
consuming micrography analysis on many planes, they managed to
give a 3D reconstruction of clusters resulting from stirring at
steady state. They gave an estimation of the entrapped liquid
fraction and plotted the measured viscosity versus EVF for
slurries solidified under different shear rates. All the
experimental points were scaled back on a unique master curve.

The critical point of this analysis is the estimation of the
structure dependent EVF, as a function of the shear rate. In the
following section, we propose an approach predicting the structure
evolution of a semi-solid slurry submitted to shear.

\section{The steady state of sheared suspension: competition between
aggregation and break-up}\label{steadysec}

When a semi-solid slurry is left at rest, two different mechanisms
occurs:
\begin{itemize}
\item Oswald ripening \cite{Wan90} and spheroidization tend
to minimize the solid/liquid surface energy by narrowing the size
distribution of aggregates, and smoothing the solid particles
surfaces. These phenomena are diffusion limited with a
characteristic length of the order of the individual particles
radius $a_0$. A characteristic diffusion time could be
approximated by: $a_0^2/D \approx 2$~s, where $D \approx 5 \times
10^{-9}$ m$^2$s$^{-1}$ is the diffusion coefficient of the solute
in the liquid and $a_0 \approx 100~\mu$m.
\item Particles aggregation occurs to lower the liquid/solid
surface energy. The driving force of aggregation is the difference
between twice the solid/liquid surface energy and the solid/solid
surface energy: $\Delta \sigma = 2 \sigma_{SL} - \sigma_b$. The
aggregation kinetic is limited by the collision frequency which is
estimated in \cite{mason57} for a sheared suspension of hard
spheres as: $f_c=(8 / \pi) \times \Phi \dot{\gamma}$. In the
intermediate region of figure~\ref{structurefig}, the
characteristic time between two collisions $(1/f_c)$ ranges from
1~s to 1~ms when $\dot{\gamma}$ ranges from 10 to 1000~s$^{-1}$.
\end{itemize}

In the present paper, we are interested in a shear rate domain
where the kinetic of aggregation is much more rapid than the
kinetic of Oswald ripening (see above). We will then neglect
Oswald ripening. Indeed, in a relevant time for diffusion limited
mechanisms, aggregation induces more drastic changes in the
microstructure than Oswald ripening.

The aggregation mechanism is collision induced. We are interested
in the physical mechanisms taking place during the contact between
two particles. Two spherical objects of radius $a_0$ aiming at one
another in a shear flow with shear rate $\dot{\gamma}$ will meet
during a mean contact time $\tau_c$ approximated by Adler
\cite{Adler94} for the limit of small solid fraction:

\begin{equation}\label{tauceq}
  \tau_c=\frac{5}{2} \dot{\gamma}^{-1}
\end{equation}

\begin{figure}
  \centering
  \includegraphics[width=\textwidth]{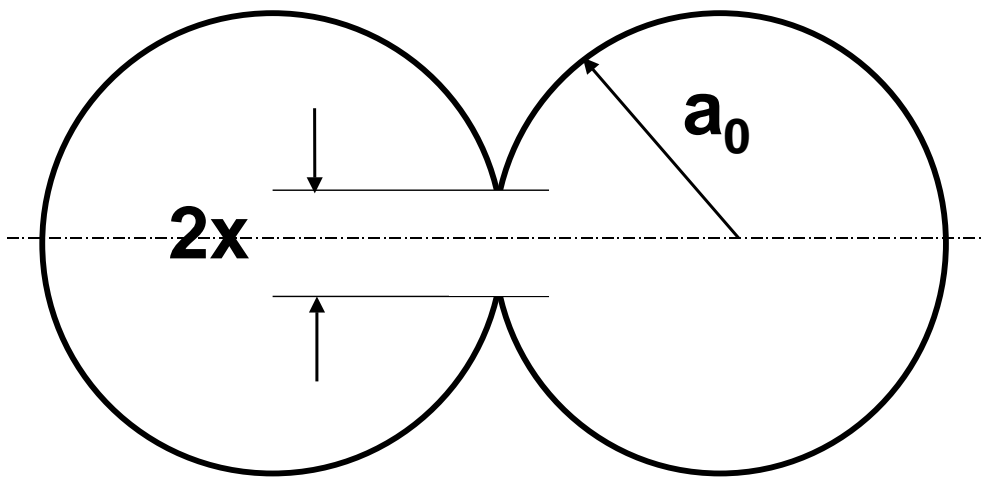}
  \caption{Neck growth between two solid particles.}\label{neckfig}
\end{figure}

During the contact, if a favorable crystallographic orientation is
encountered, a rigid neck will be built between the two particles
(see figure \ref{neckfig}). We now use a result derived in
reference \cite{Martin94} concerning the growth rate of a neck
between two solid spherical particles of radius $a_0$ (see
figure~\ref{neckfig}). If $x$ is the neck radius and $\tau_f$ the
neck formation time after the particles collision, $\tau_f$
follows:

\begin{equation}\label{xeq}
  \tau_f = \frac{1}{5} A \left( \frac{x}{a_0} \right)^5 a_0^3
\end{equation}

where $A$ is a constant dependent on thermo-physical properties of
the material (see reference \cite{Barbé00} for more details).
Assuming the neck builds up during the contact time $\tau_c$, we
have $\tau_c = \tau_f$, leading to an estimation of the neck size:

\begin{equation}\label{neckeq}
  x = \left( \frac{25}{2A} \right)^\frac{1}{5} \dot{\gamma}^{-\frac{1}{5}}
    a_0^\frac{2}{5}
\end{equation}

The shear will tend to break the aggregates during collision. As
collision of two clusters  of size $\Delta L$ occurs their mean
relative velocity is $\Delta L/2 \times \dot{\gamma}$. Thus, the
kinetic energy $E_c$ available for rupture is:

\begin{equation}\label{kineticeq}
  E_c = \frac{1}{2} \rho V \left(\dot{\gamma} \frac{\Delta L}{2}\right)^2
\end{equation}

$\rho$, $V$, $\dot{\gamma}$ and $\Delta L$ are the mass density,
the volume, the shear rate and the size of the cluster in the
normal direction of the rupture plane, respectively. $E_c$ has to
be compared with the energetic cost of the plastic rupture
(classically considered to be 1000 times larger than the fragile
rupture energy \cite{Ashby96}):

\begin{equation}\label{ruptureq}
    E_r= n \times \pi x^2 \times 1000 \Delta \sigma
\end{equation}

$n$, and $\Delta \sigma$  being the number of broken necks of size
$x$ (defined in equation \ref{neckeq}) in the fracture plane, and
the energy cost due to liquid / solid surface creation,
respectively.

If $\dot{\gamma}$ is small, the contact time $\tau_c$ is long and
necks will be large. Moreover, kinetic energy will be low
preventing from any break-up of the resulting cluster. This
situation will lead to the formation of a rigid 3D network. For
large $\dot{\gamma}$ the contact time is short and the kinetic
energy is large enough to prevent any aggregation. This leads to a
dispersed suspension. As both aggregation and break-up are
collision induced, the collision frequency influences only the
kinetic of aggregation~/~break-up, but it does not influence the
equilibrium between these two mechanisms.

In the present study, we deal with "liquid driven deformation".
This means that the shear rate is sufficiently high to allow the
competition between aggregation and break-up, the deformation
being allowed by the free liquid between the clusters.

From these simple considerations the probabilities of aggregation
and break-up between particles and clusters of particles will be
derived and used to simulate cluster dynamics.

\section{Lattice model simulation of cluster structure}\label{simulsec}

In a real system containing $\approx 10^5$ particles, one would
have to treat $\approx 10^3$ clusters in order to be
representative of the distribution. This is untractable with
standard computer power. We are interested here in steady state
interaction, which means a dynamical equilibrium between cluster
aggregation and cluster fracture. We assume a sort of "ergodic"
hypothesis, namely that in steady state, the size distribution of
a population of clusters at a given time, is equivalent with the
size distribution of a given cluster along time. Our main
simplifying approach based on the above hypothesis is in
considering only one representative cluster that can either
aggregate with its copy, or can be broken in a random plane. The
time evolution of this cluster will approximate the ensemble
properties of the equilibrium semi-solid slurry.

The cluster is made up of a cubic arrangement of connected spheres
(elementary particles of radius $a_0$). It is stored in a 3D
matrix (cubic lattice), where 0 stands for free liquid, 1 stands
for solid and 2 stands for entrapped liquid. The size of the
matrix unit cell ($2a_0$) is chosen so that the elementary
particle is inscribed into this cubic unit cell.

\subsubsection*{Aggregation}

The aggregation probability $P_a$ per collision is assumed to be
the probability $q_e$ for two particles to encounter with a
favorable crystallographic orientation.  As in reference
\cite{Martin94}, we consider that only low angle grain boundary
($\le$~0.25~rad) give rise to a non wetted grain boundary, leading
to: $q_e=0.02$. This condition is equivalent to:

\begin{equation}\label{stickprobaeq}
  P_a = q_e
\end{equation}

\subsubsection*{Break-up}
The rupture probability per collision $P_r$ should be a monotonous
increasing function of the ratio of the available kinetic energy
over the energetic cost of the rupture ($E_c/E_r$). At zero
kinetic energy it should be zero, and for $E_c=E_r$ it should
converge to $1$. The most simple form satisfying the above
criterion is the simple $P_r=E_c/E_r$ approximation. A fracture
plane is randomly chosen. The differential kinetic energy
introduced in \cite{Barbé00} is given by equation~\ref{kineticeq}.
The volume $V$ of the cluster is taken as $(e+s) (2a_0)^3$ with
$e$ and $s$ being the number of entrapped liquid voxels and the
number of solid particles in the cluster, respectively. The
energetic cost of the rupture is given by equation~\ref{ruptureq}.
We then have the rupture probability per collision:

\begin{equation}\label{breakprobaeq}
  P_r = \frac{E_c}{E_r}
  = \frac{2}{125} \frac{\rho a_0^{\frac{21}{5}}}{\pi \Delta \sigma}
  \left( \frac{25}{2A} \right)^{-\frac{2}{5}}
  \times \frac{(e+s) \left( \frac{\Delta L}{4 a_0} \right)^2
  \dot{\gamma}^\frac{12}{5}}{n}
\end{equation}

The first group of terms in \ref{breakprobaeq} depends on material
properties ($\rho$, $\Delta \sigma$, $A$) and the initial state
($a_0$), whereas the second member is calculated at each step of
the simulation.

During cluster/cluster collisions, aggregation or rupture may
occur. However, most of the time, neither aggregation nor break-up
occur and the two clusters are left as before. As we deal with the
steady state there is no need to consider these collisions with no
effect. The only parameter is then the ratio $P_a/P_r$.

\begin{equation}\label{simconsteq}
\begin{array}{l}
  \frac{P_a}{P_r} = K \times \frac{n}
  {(e+s) \left( \frac{\Delta L}{4 a_0} \right)^2
  \dot{\gamma}^\frac{12}{5}}\\
  K = \frac{125 \pi}{2} \frac{q_e \Delta \sigma}{\rho a_0^{\frac{21}{5}}}
  \left( \frac{25}{2A} \right)^{\frac{2}{5}}
  \end{array}
\end{equation}

Through equation~\ref{simconsteq}, $K$ incorporates the material
properties and the initial state. For an Al-6.5wt\%Si alloy with a
50~$\mu$m globular structure, $K = 5\times10^{8}$~s$^{-12/5}$ (see
Table~\ref{symboltab}).

For each step, the one-cluster algorithm randomly selects a
rupture plane and calculates the break-up probability
(aggregation probability is fixed to $q_e$). A number is randomly
generated between 0 and $P_a + P_r$:
\begin{itemize}
\item If it ranges between 0 and $P_r$ fracture is processed by
deleting all the solid particles belonging to the rupture plane.
One of the resulting clusters is randomly selected and stored in
the matrix $M$.
\item Else, the cluster is duplicated in a matrix $N$, randomly
rotated, and the two matrices, $M$ and $N$, are then aggregated
allowing interpenetration. The resulting cluster is stored in the
matrix $M$.
\end{itemize}

\begin{figure}
  \centering
  \includegraphics[width=\textwidth]{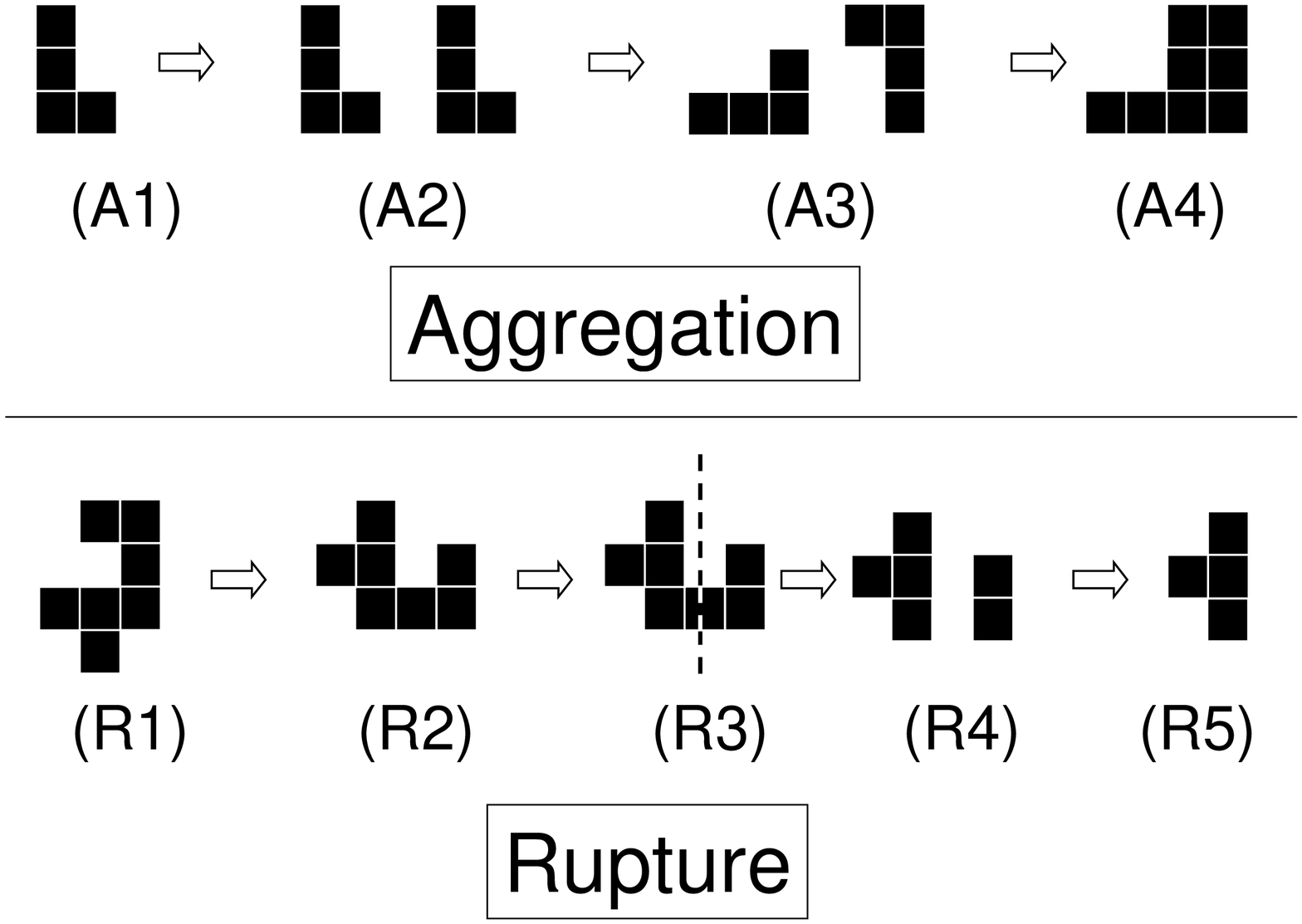}
  \caption{Aggregation process: (A1) Initial cluster -
  (A2) Initial cluster is duplicated - (A3) Random rotation -
  (A4) Sticking of the two clusters. Rupture process:
  (R1) Initial cluster - (R2) Random rotation - (R3) Random selection
  of a fracture plane - (R4) Erasing the particles in the fracture plane -
  (R5) Random selection of the resulting cluster. Note that the fracture
  probability depends on the number of neck to be broken in the
  fracture plane.}\label{aggregruptfig}
\end{figure}

\begin{figure}
  \centering
  \includegraphics[width=\textwidth]{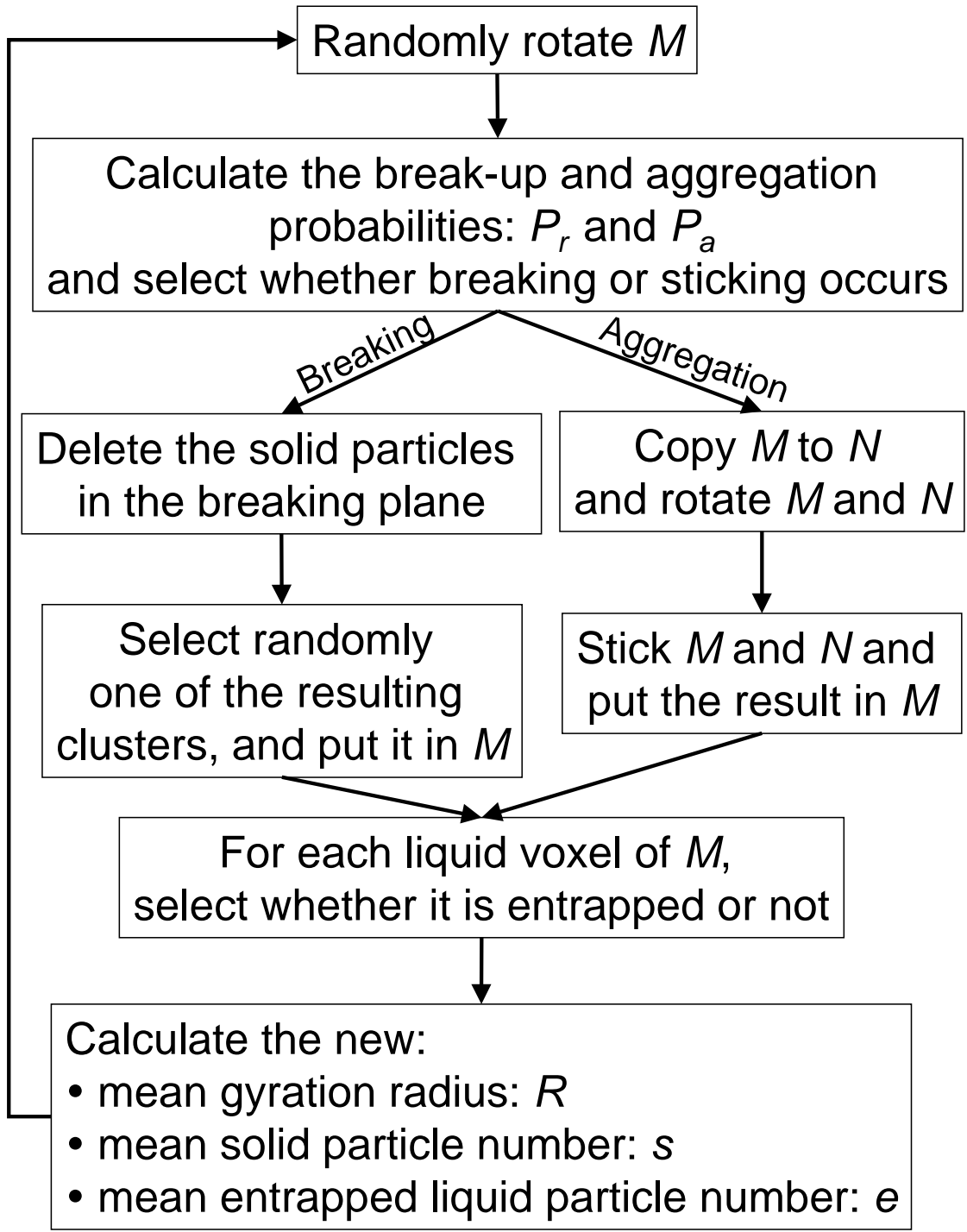}
  \caption{Simulation algorithm: the cluster is stored in $M$ and $N$: 3D
  matrices.}\label{algofig}
\end{figure}

Aggregation and rupture mechanisms are illustrated in
figure~\ref{aggregruptfig} and a simplified algorithm of the
numerical simulation is shown in figure \ref{algofig}.

The number of solid particles $s$, the number of entrapped liquid
voxels $e$ and the gyration radius $R$ are extracted from the
matrix $M$, leading to the calculation of the effective volume
fraction. The gyration radius is defined as the mean distance
between the cluster center of mass and all its solid particles.
Averages ($\overline{R}$, $\overline{s}$, $\overline{e}$) are
calculated over the total step number. Since the mean "life-time"
between two collisions is assumed to be independent of the cluster
radius, there is no need to weight the averages.

\begin{figure}
  \centering
  \includegraphics[width=\textwidth]{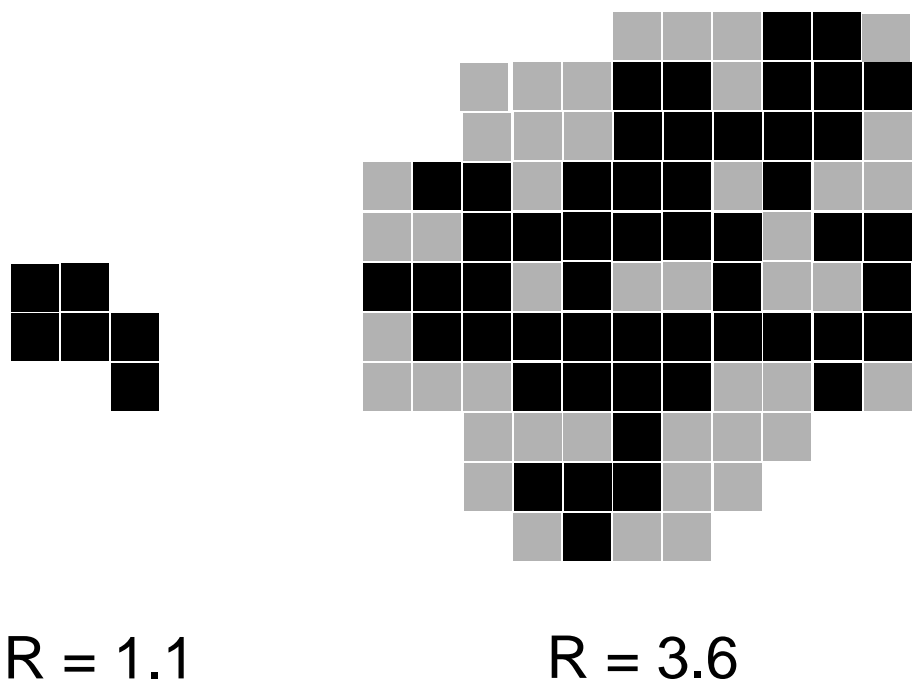}
  \caption{Representation of the entrapped liquid (grey) for a 2D Cluster:
  The liquid is entrapped if three or four of the  principal directions
  point to any solid particle (black). In the case of two directions pointing
  to solid particles, the mean distance to solid is compared with the
  cluster gyration radius.}\label{entrapprocfig}
\end{figure}

One of the crucial point of this simulation is the evaluation of
the entrapped liquid in the cluster. Indeed, this effect governs
the effective volume fraction $\Phi_{eff}$ that will be introduced
in equation~\ref{kriegermod} to predict the viscosity. It is
calculated at each step and defined as follows: a liquid voxel is
entrapped if 4 or more of its 6 principal directions (cubic
lattice) $(\overrightarrow{Ox},-\overrightarrow{Ox},
\overrightarrow{Oy}, -\overrightarrow{Oy}, \overrightarrow{Oz},
-\overrightarrow{Oz})$ hits any solid particle of the cluster. If
only 3 directions out of the 6 point to a solid particle, the
liquid is entrapped if the averaged distance between the
considered liquid voxel and the 3 intersected solid particles is
smaller than the cluster gyration radius $R$. In other terms, if
$d_1$, $d_2$ and $d_3$ are the distances between the liquid voxel
and the nearest solid particle in the 3 considered directions, the
liquid is entrapped if $d_1^2+d_2^2+d_3^2<R^2$. For a better
understanding, this procedure is visualized in
figure~\ref{entrapprocfig} for the 2D case.

\section{Numerical validation:}\label{numsec}

\begin{figure}
  \centering
  \includegraphics[width=\textwidth]{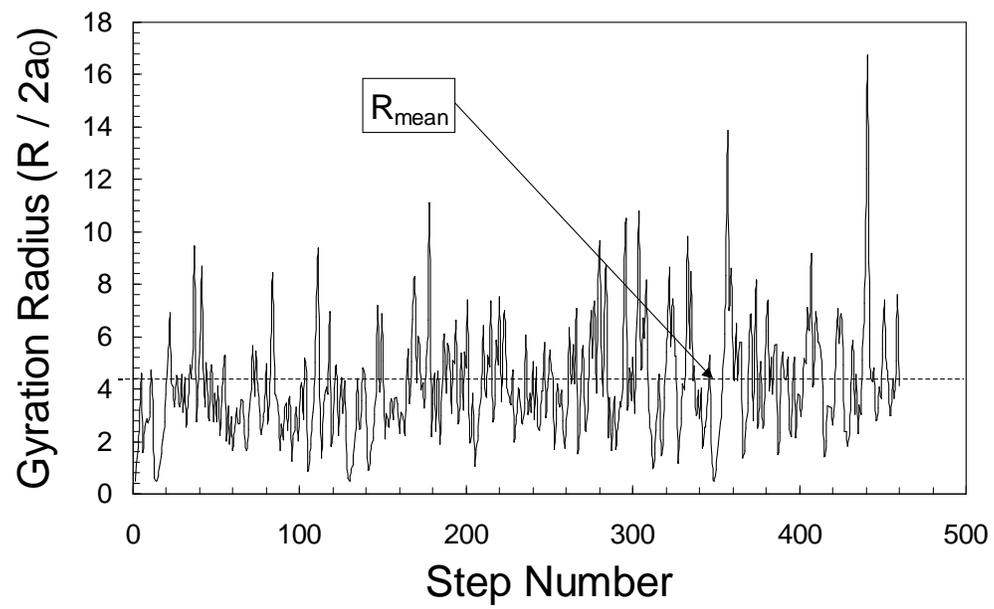}
  \caption{Time evolution of the cluster gyration radius. $R$ fluctuates a lot,
  but its time evolution is assumed to be representative of the spatial
  distribution of the clusters.}\label{radiusfig}
\end{figure}

\begin{figure}
  \centering
  \includegraphics[width=\textwidth]{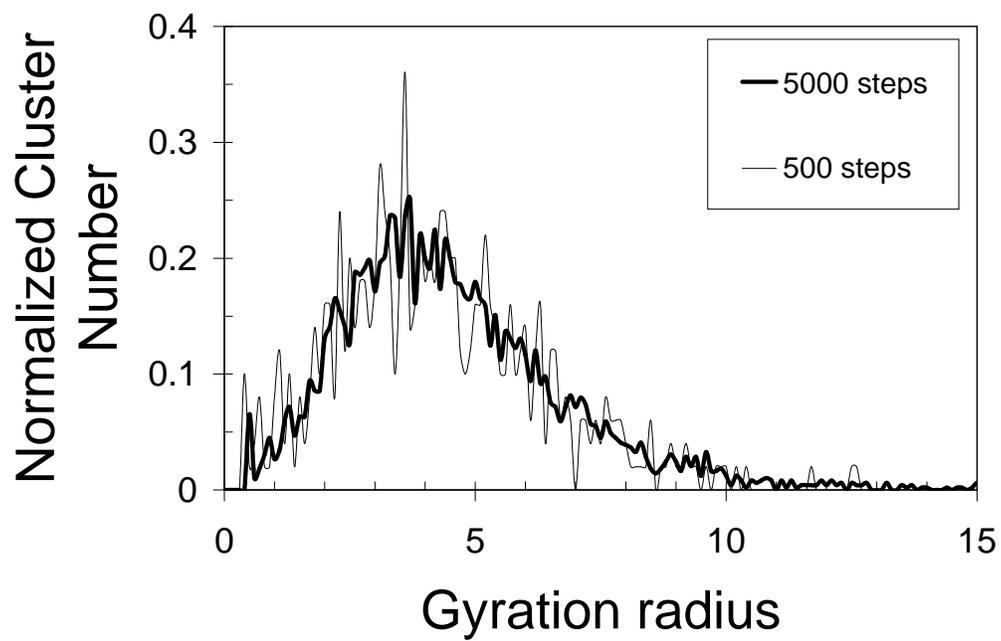}
  \caption{distribution of $R$ after various computation steps.
  Steady state is reached when
distribution fluctuations are less than
10\%.}\label{steadydistribfig}
\end{figure}

The simulation starts with a cluster constituted of one solid
particle. It usually growths until it reaches a large enough size
to be broken. A typical time evolution of the cluster gyration
radius $R$ is shown in figure \ref{radiusfig} for a shear rate of
500 s$^{-1}$. Although the gyration radius $R$ fluctuates strongly
in time, the distribution of $R$ is stable. The numerical
convergence is reached when fluctuations of the distribution are
less than 10\%. Figure~\ref{steadydistribfig} displays the cluster
gyration radius distribution after various computation steps.
Steady state is reached after $\approx 5000$ steps.

\section{Results}\label{resultsec}

The result of the simulation will be presented as follows:
\begin{enumerate}
\item 3D cluster shape at different shear rates will be depicted.
\item The size distribution of clusters for different shear rates
will be plotted and compared with an analytical model available in
the literature.
\item  The $\dot{\gamma}$ dependence of the mean radius
$\overline{R}$ of the clusters will be discussed.
\item The parameter $\Phi_{eff} / \Phi$
will be computed as a function of the shear rate giving a more
precise structure map than the one given in
figure~\ref{structurefig}.
\item Finally, the viscosity $\eta$ will be given as a function of
the shear rate and the solid fraction.
\end{enumerate}

\subsubsection*{Cluster shape}

\begin{figure}
  \centering
  \includegraphics[width=\textwidth]{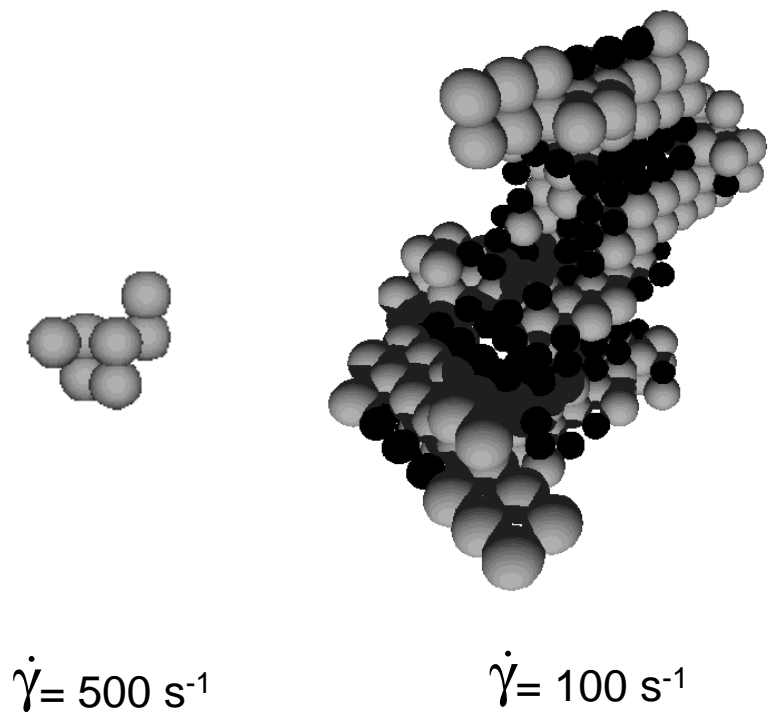}
  \caption{Clusters for two different shear rates. Liquid is transparent,
  solid is grey, and entrapped liquid is black. Larger cluster corresponds
  clearly to higher entrapped liquid fraction.}\label{clusterfig}
\end{figure}

Figure \ref{clusterfig} shows a cluster of radius $R$ with its
entrapped liquid for two different shear rates. For small
$\dot{\gamma}$, aggregation is dominant leading to a more open
structure entrapping more liquid.

\subsubsection*{Cluster size distribution}

\begin{figure}
  \centering
  \includegraphics[width=\textwidth]{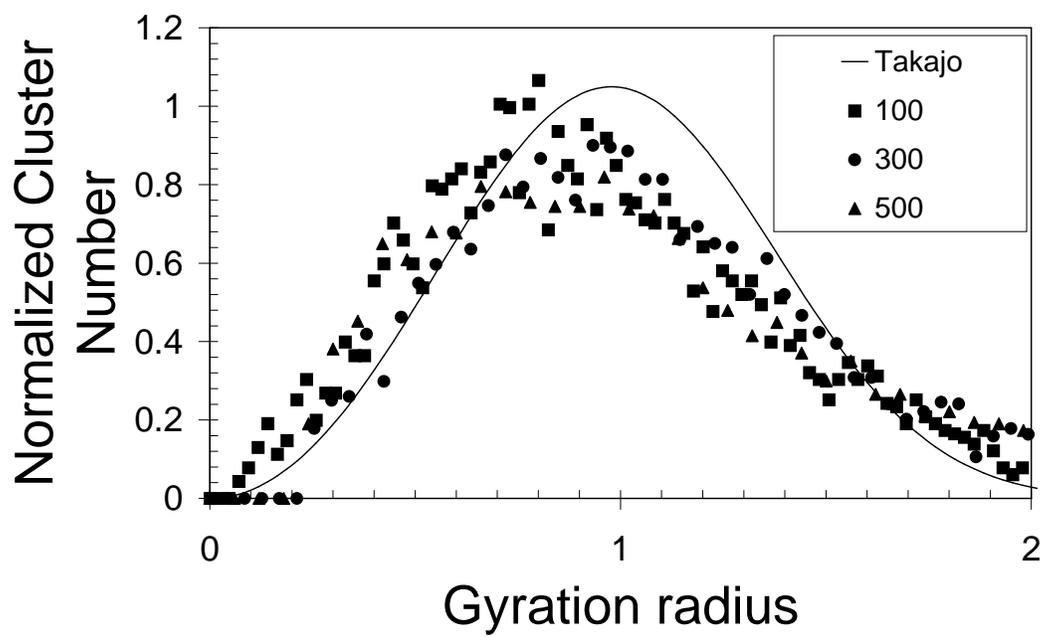}
  \caption{Gyration radius distribution for $5000$ simulation steps at
  various shear rates compared with the mathematical model of
  Takajo \cite{Takajo84}
  }\label{distribfig}
\end{figure}

Figure \ref{distribfig} exhibits normalized radius distributions
for different shear rates. Size distribution of particles can be
accurately described by a "log-normal" law, frequently encountered
in fragmentation problems (see reference \cite{Reed88}). Since
these experiments are ensemble distributions, whereas our
simulation deals with time evolution, this similarity gives some
confidence into the used "ergodic"-type hypothesis.

The normalization brings the simulated distribution for various
shear rates to a single master curve. The size dispersion is then
proportional to the mean size of the clusters: lower is the shear
rate, larger is the mean cluster radius and larger is the size
dispersion. For low shear rate, the cluster can explore either
small and large sizes, whereas, for high shear rate, only small
sizes are possible.

In figure \ref{distribfig}, one can compare the simulated
distribution with the mathematical model of Takajo \cite{Takajo84}
often used as a model for aggregation processes. This model based
on homogeneous coalescence frequency gives a normalized size
distribution at steady state. These two different approaches lead
to comparable profiles, giving some confidence into the general
basis of our simulation.

\subsubsection*{Mean cluster radius {\it vs.} shear rate}

\begin{figure}
  \centering
  \includegraphics[width=\textwidth]{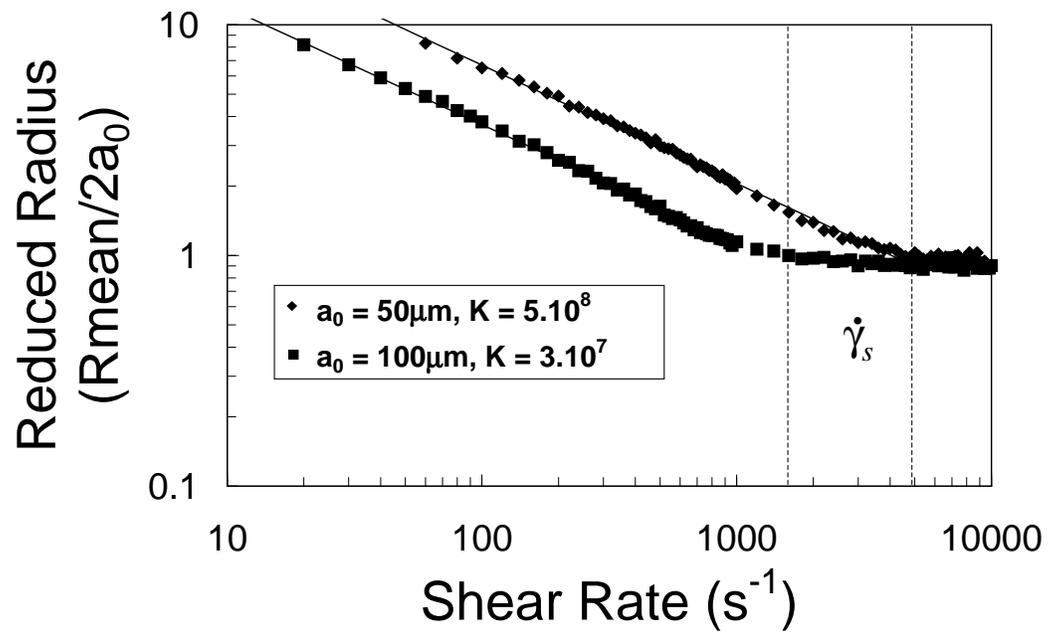}
  \caption{The $\dot{\gamma}$ dependence of the mean gyration radius.
  For $\dot{\gamma}$ higher than $\dot{\gamma_s}$,
  the shear breaks the clusters into individual particles.}
\label{rgammapfig}
\end{figure}

The obtained mean gyration radius values are plotted as a function
of the shear rate on figure~\ref{rgammapfig} The mean gyration
radius $\overline{R}$ decrease with increasing shear rate
following a power law $\overline{R} \propto \dot{\gamma}^{-0.7}$.
This dependence is comparable with the power law estimated for the
compact cluster scenario \cite{Barbé00}: $R \propto
\dot{\gamma}^{-0.6}$. It can be observed that, above a given
$\dot{\gamma}$ only elementary particles remain, hence there is no
more evolution of $\overline{R}$.

As the physical parameter $K$ depends strongly on the size of the
individual particles $a_0$ (see equation~\ref{simconsteq}),
computations are presented for two different reasonable values of
$a_0$. For the smaller elementary particle size, cluster are
larger at some given shear rate and solid fraction.

\subsubsection*{Effective volume fraction {\it vs.} shear rate}

\begin{figure}
  \centering
  \includegraphics[width=\textwidth]{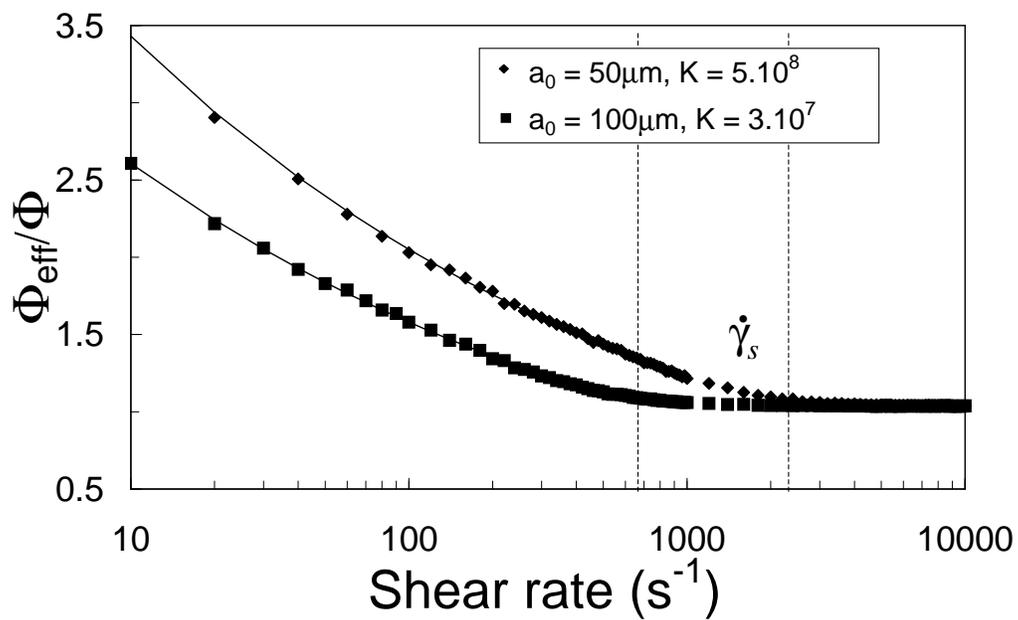}
  \caption{The ratio $\Phi_{eff}/\Phi$ as a function of $\dot{\gamma}$
  for two values of the simulation constant corresponding to
  two initial particle sizes. For $\dot{\gamma}$ higher then $\dot{\gamma_s}$,
  the structure is too small to trap any liquid.}
  \label{phiefffig}
\end{figure}

Figure \ref{phiefffig} displays the calculated ratio
$\Phi_{eff}/\Phi$ as a function of $\dot{\gamma}$ for two
different elementary particle sizes. For high $\dot{\gamma}$,
(high enough to break all the necks), there is no entrapped liquid
($\Phi_{eff}/\Phi$ = 1). For small $\dot{\gamma}$, more connected
structures leads to high fraction of entrapped liquid
($\Phi_{eff}/\Phi \approx 2 $).

\begin{figure}
  \centering
  \includegraphics[width=\textwidth]{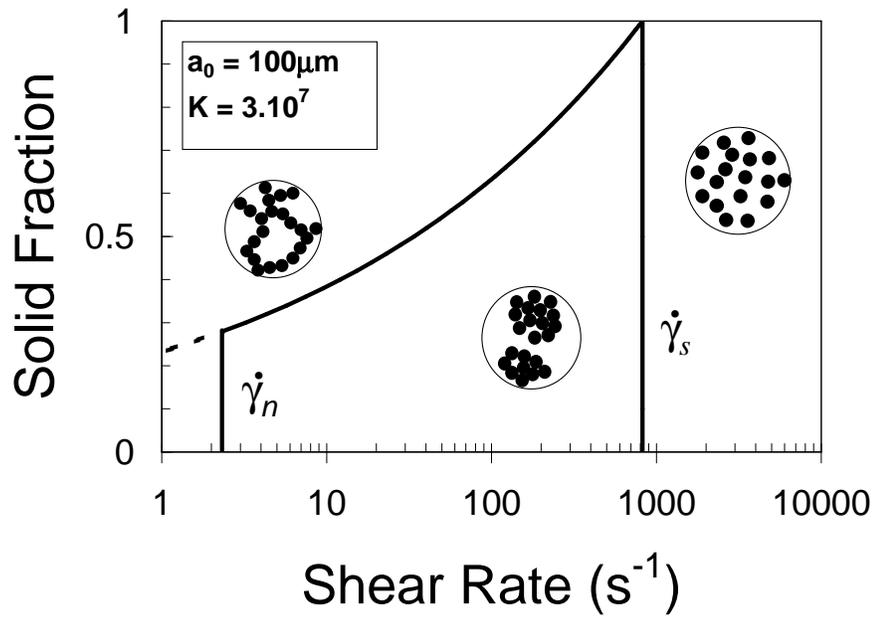}
  \caption{Suspension structure depending on $\dot{\gamma}$ and
  $\Phi$.  $\dot{\gamma_n}$ is now $\Phi$ dependent, except for
  low solid fraction where gelation occurs when a cluster reach
  the size of the apparatus gap (left vertical line).}
  \label{newstructurefig}
\end{figure}

In figure~\ref{newstructurefig} we can identify the three
different domains introduced in section~\ref{rheosec}
(figure~\ref{structurefig}).

\begin{itemize}
\item For high $\dot{\gamma}$ values ($\geq \dot{\gamma_s}$), the dispersed
suspension-like behavior is observed. $\dot{\gamma_s}$ is reached
when the mean radius of the clusters is equal to the radius of the
elementary solid particles, that is to say when the effective
volume fraction is equal to $\Phi$. Note that an approximation of
$\dot{\gamma_s}$ can be obtained equaling $P_a$ and $P_r$ of
equation \ref{simconsteq}. We find for Al-6.5wt\%Si with initial
particles size $a_0 = 100~\mu m$: $\dot{\gamma_s}=1300~s^{-1}$.
This value is clearly in accordance with the transition domain of
figure \ref{phiefffig}.
\item For low $\dot{\gamma}$ values ($\leq~\dot{\gamma_n}$) a rigid network
sets in. The estimation of $\dot{\gamma_n}$ depends on the
gelation mechanism:

\begin{itemize}
\item experimentally, the apparent rigid network will appear when
the mean radius of the clusters reaches the characteristic size of
the measurement apparatus (Couette Rheometer for example).
\item in theory, the apparent rigid network will appear when
$EVF=\Phi_M$: the whole volume of the sample is invaded by a
unique cluster.
\end{itemize}

$\dot{\gamma_n}$ is found to be $\Phi$ dependent. We can observe
that the simulation leads to a more realistic description of the
structure(figure~\ref{newstructurefig}) than in
figure~\ref{structurefig}. Indeed, lower is the solid fraction,
higher is the liquid fraction involved in the hydrodynamic flow
for some given $\dot{\gamma}$ and aggregation state characterized
by $e$ and $s$.
\item In the intermediate range we have a suspension
of clusters with decreasing size and entrapped liquid fraction as
$\dot{\gamma}$ increase.
\end{itemize}

\subsubsection*{Viscosity {\it vs.} shear rate}

For the calculation of the viscosity, the entrapped liquid is
taken into account trough (a) the mean number of entrapped liquid
voxels $\overline{e}$, and (b) the amount of entrapped liquid
within a cubic voxel containing a solid sphere, which depends on
the mean number $n_n$ of solid neighbors of a particle over the 26
possible neighbors in the cubic lattice. The liquid fraction
within a unit cell is 0.42 (difference of volume between the cube
and its inscribed sphere). The viscosity follows:

\begin{equation}
\label{viscoeq} \begin{array}{l}
    \eta = \eta _0 \left( 1-
        \frac{\Phi_{eff}}{\Phi_M} \right) ^{-2.5 \Phi_M} \\
    \Phi_{eff} = \Phi \left( 1+ \frac{\overline{e}}{\overline{s}} + \frac{(0.42)
    \frac{n_n}{26}}{s} \right)
    \end{array}
\end{equation}

where $\overline{e}$ and $\overline{s}$ are the mean number of
entrapped liquid sites and the mean number of solid sites in the
cluster, respectively. With a suspension of hard spheres, $\Phi_M$
would have been 0.65, but as we deal with polydisperse and
deformable solid particles, it is chosen to be 1. Indeed, the
liquid phase has been proven to remain connected for solid
fraction up to 0.8.

\begin{figure}
  \centering
  \includegraphics[width=\textwidth]{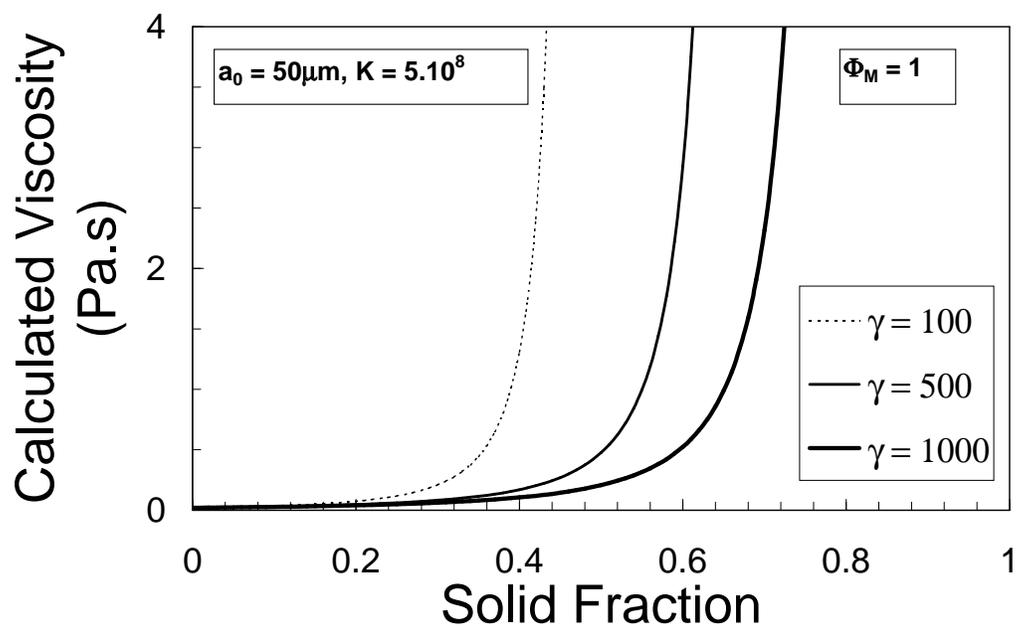}
  \caption{Viscosity as a function of the solid fraction for
different shear rate. Higher is the shear rate, higher is the
solid fraction leading to gellation.}\label{viscofig}
\end{figure}

These equations lead to figure \ref{viscofig} where viscosity is
plotted versus the solid fraction. The viscosity exhibits the
typical profile of a semi-solid mixture \cite{Joly76}. The
viscosity increases with the solid fraction until the solid
fraction reaches a threshold value where the suspension behaves
more like a solid than a liquid. Higher is the shear rate, higher
is this threshold value.

\section{Comparison with real microstructure and experimental
viscosity measurement}\label{compasec}

\subsection{3D microstructure visualization}

\begin{figure}
  \centering
  \includegraphics[width=\textwidth]{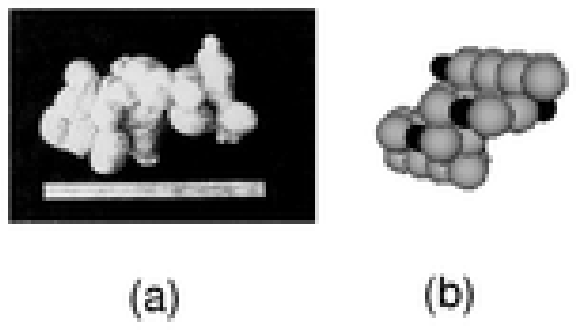}
  \caption{Experimental and simulated cluster structure taken from
  an Al-6.5wt\%Si semi-solid slurry sheared at 900 $s^{-1}$. (a)
  3D reconstruction extracted from reference \cite{Ito92}
  (b) Simulated with $K = 5 \times 10^8$}
  \label{agglo3dfig}
\end{figure}

The advantage of such a simulation is the ability of viewing 3D
structures. It is of interest to compare them with real structures
resulting from serial cutting and micrography analysis made by Ito
\cite{Ito92}. Figure \ref{agglo3dfig} shows experimental and
simulated 3D cluster structures. The physical parameter $K$ (see
equation \ref{simconsteq}) has been calculated for the
Al-6.5wt\%Si alloy with a elementary particle radius $a_0 = 50
\mu$m. Note that the cluster exhibited by Ito is not
representative of the population. Thus, we extracted from the
simulation a cluster with comparable size.

\subsection{Viscosity measurement}

Comparison with experimental viscosity measurement is not an easy
task. Viscosity measurements are usually performed after
solidification under shear rate \cite{Ito92}. Both the
solidification and the aggregation processes takes place
simultaneously.

\begin{figure}
  \centering
  \includegraphics[width=\textwidth]{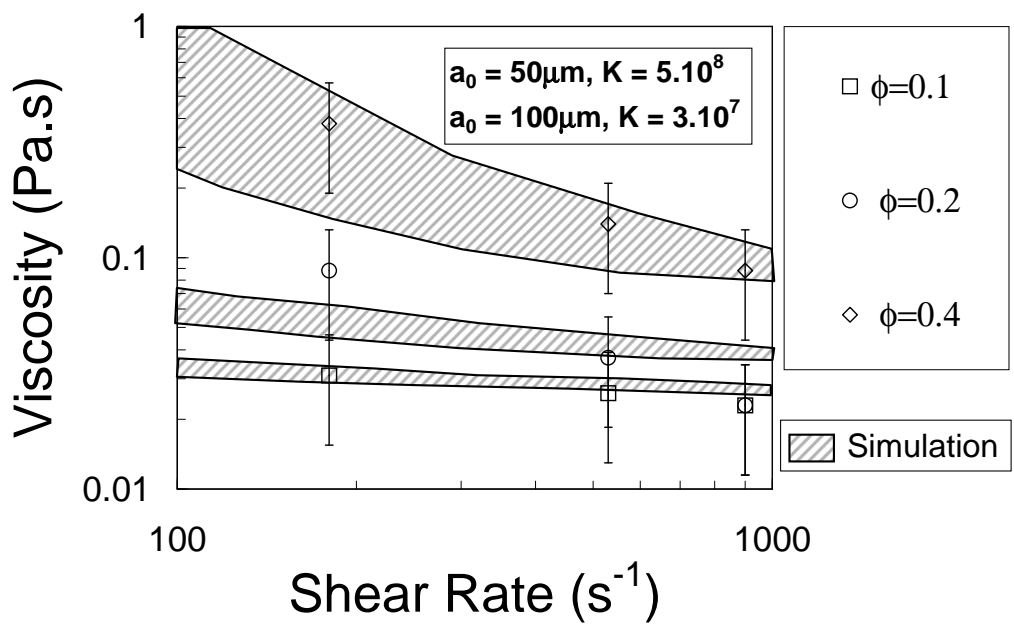}
  \caption{Comparison between simulation results and experimental
  viscosity measurement of an
  $Al-6.5wt\%Si$ alloy \cite{Ito92} as a function of the
  shear rate and solid fraction. Simulation results are spread over a
  domain limited by two reasonable values of $a_0$. This domain contains
  the experimental steady state measurements.}\label{compfig}
\end{figure}

In our approach a suspension of dispersed individual globular
particles is submitted to shear until it reaches its steady state
of aggregation. This procedure decorrelates solidification and
aggregation. However, the solidification process is assumed to
have a slighter effect than the shear on both the structure and
the viscosity. We then compare the simulation results with
experimental viscosity measurements performed on Al-6.5wt\%Si
alloys partially solidified under stirring \cite{Ito92}. Those
experimental results were obtained with Couette rheometer and the
measurements were performed at steady state: when the shear stress
is stable over time. This experimental procedure allows us to
consider that the aggregation degree of the microstructure
corresponds to the steady aggregation degree. The good agreement
between simulation results and experimental results plotted in
figure~\ref{compfig} tends to validate both the former assumption
and the basis of the simulation.

\subsection{Particle size distribution}

\begin{figure}
  \centering
  \includegraphics[width=\textwidth]{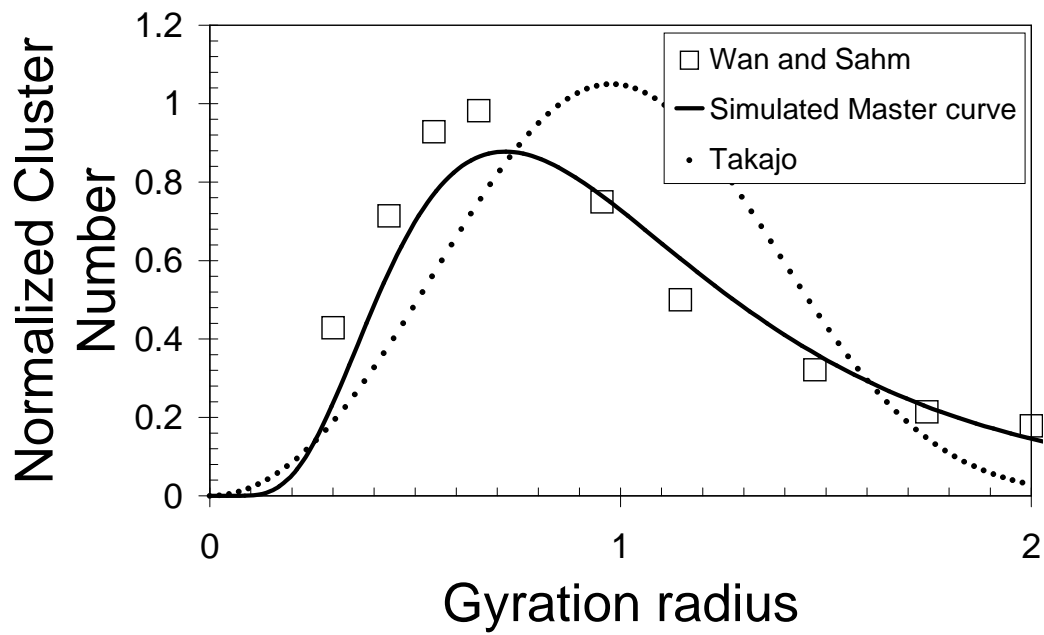}
  \caption{Normalized cluster size distribution: comparison
  between experimental microstructure analysis of Wan \cite{Wan90},
  and the present simulation. Result of Takajo~\cite{Takajo84} are plotted
  for comparison. The simulated distribution seem to fit
  accurately the real distribution profile.}\label{normdistribfig}
\end{figure}

In figure~\ref{normdistribfig}, we plotted the simulated
distribution master curve (see section~\ref{resultsec}) and
experimental normalized size distributions resulting from
microstructure analysis \cite{Wan90}. The master curve seem to fit
accurately the real distribution profile.

\section{Conclusion}\label{conclusec}

The structure of semi-solid slurries results from the competition
between aggregation due to surface forces and break-up induced by
the shear field. Aggregation probability is set to be constant.
Break-up probability is then set to be proportional to the ratio
of the available kinetic energy over the rupture energy. Using a
one-cluster algorithm, we supposed that the time evolution of a
single cluster approximates the ensemble distribution of a
population of clusters ("ergodic" hypothesis). Simulation of the
time evolution of our representative cluster describes accurately
the structure of the slurry with no adjustable parameter and
estimates successfully the viscosity as a function of the applied
shear rate and the volumic solid fraction. The present work allows
a better understanding of the structure evolution of sheared
semi-solid metallic alloys in the ($\Phi, \dot{\gamma}$) plane. In
a future work, ripening could be also taken into account through a
reinforcement of the necks along time after the particles
encounter.

The present approach could be easily transferable to any type of
suspension as far as the interacting forces between solid
particles can be estimated.

The evaluation of the entrapped liquid fraction could be performed
using a general dimension code for computing convex hulls.
"QuickHulls" from the Geometry Center, University of Minnesota,
USA, could provide us with a powerful tool.

\section*{Acknowledgment}
We are grateful to C. Martin, M. Papoular, M. Su\'ery and A.
Zavaliangos, for useful discussions. This work has been partly
funded by Electricit\'e De France (EDF) and CEA. One of us
(Michel Perez) is supported by a MENRT grant and partial
financial support from R\'egion Rh\^one-Alpes is gratefully
acknowledged.

\bibliography{biblio}
\newpage

\setlongtables
\begin{longtable}[c]{lp{13.5cm}}
   \hline
    Symbol  &   Meaning\\
    \hline \endfirsthead
    \hline
    Symbol  &   Meaning\\
    \hline \endhead

    $A$     &  Constant depending on thermophysical parameters
                ($1.7 \times 10^{16}~s~m^{-3}$) \cite{Barbé00}\\
    $a_0$   &  Initial particles Radius \\
    $D$     &  Diffusion coefficient of the solute in the liquid
                ($5 \times 10^{-9}$ m$^2$s$^{-1}$)\\
    $d_1, d_2, d_3$   &  Distance between a liquid voxel and the
              nearest solid particle in the directions
              $(\overrightarrow{Ox}$), $(\overrightarrow{Oy}$), $(\overrightarrow{Oz})$\\
    $e$, $\overline{e}$  &  Number and mean number of entrapped liquid
                        voxels in a cluster \\
    $E_b$   &  Rupture energy of a cluster\\
    $E_c$   &  Kinetic energy of a cluster\\
    $f_c$   &  Collision frequency of clusters in a shear field\\
    $K$     &  Physical constant defined in equation \ref{simconsteq}\\
    $M, N$  &  3D Matrice used to store the cluster\\
    $n$     &  Number of necks in the fracture plane \\
    $P_r$   &  Breaking probability \\
    $P_a$   &  Aggregation probability \\
    $q_e$   &  Probability for two particles to encounter
              with favorable crystallographic orientation \\
    $R$, $\overline{R}$     &  Cluster gyration radius and cluster mean gyration radius \\
     $s$, $\overline{s}$     &  Number and mean number of solid particles in a cluster \\
    $V$     &  Volume of the cluster
               (solid + entrapped liquid)\\
    $x$     &  Neck radius\\
    $\Delta L$& Size of the cluster in the normal direction
              of the rupture plane\\
    $\Delta \sigma$&Surface Energy due to liquid / solid surface
                 creation ($0.34$ J~m$^{-2}$)\\
    $\eta$  &   Viscosity of the suspension\\
    $\eta_0$&   Viscosity of the liquid (20 mPa.s) \cite{Ito92}\\
    $\dot{\gamma}$&Shear rate\\
    $\dot{\gamma}_g$& Gelation shear rate \\
    $\dot{\gamma}_s$&Shear rate from which no aggregation occurs\\
    $\Phi$  &   Volumic solid fraction\\
    $\Phi_{eff}$&   Effective Volume Fraction (EVF)\\
    $\Phi_M$&   Maximum packing fraction\\
    $\rho$  &   Density (2350 kg~m$^{-3}$)\\
    $\sigma_{SL}$&Liquid / Solid surface energy ($0.17$ J~m$^{-2}$)\\
    $\sigma_b$& Grain boundary surface energy (negligible:
               Coincidence Site Lattice hypothesis)\\
    $\tau_c$&  Contact time between two particles
              in a shear field\\
    $\tau_f$&   Formation time of a neck between two particles\\
    \hline
\caption{Symbols and values for Al-6.5wt\%Si} \label{symboltab}
\end{longtable}

\end{document}